\documentstyle[12pt]{article}
\begin{document}
\textwidth 6.5in
\def\lsim{\:\raisebox{-0.5ex}{$\stackrel{\textstyle<}{\sim}$}\:}
\def\gsim{\:\raisebox{-0.5ex}{$\stackrel{\textstyle>}{\sim}$}\:}
\begin{flushright}
TIFR/TH/93-14
\end{flushright}
\bigskip
\bigskip
\bigskip
\begin{center}
\large {\bf R-PARITY VIOLATING SUSY MODEL\footnote{Invited Talk at the X
Symposium on High Energy Physics, TIFR, Bombay, 26-31 December 1992 -- To
be Published in Pramana, Proceedings Suppliment (1993)}} \\
\bigskip
\bigskip
{\bf D.P. Roy} \\
Theoretical Physics Group \\
Tata Institute of Fundamental Research \\
Homi Bhabha Road, Bombay 400 005, India \\
\bigskip
\bigskip
{\bf Abstract} \\
\end{center}
\bigskip

This is a phenomenological review of $R$ parity violating SUSY models,
with particular emphasis on explicit $R$ parity violation.

\newpage

\noindent {\underbar{\bf Introduction}} : \\

\smallskip

The Minimal Supersymmetric Standard Model (MSSM) contains the particles of
the Standard Model (SM) with two Higgs doublets as shown in the first row
below and their supersymmetric partners as shown in the second [1].  They
can be combined to form the corresponding super multiplets as shown in the
third row.
$$
\begin{array}{l}
\pmatrix{\ell_i,\bar e_i,q_i,\bar u_i,\bar d_i,h_1,h_2,g,W,Z,\gamma \\
\tilde \ell_i,\tilde {\bar e}_i,\tilde q_i,\tilde {\bar u}_i,\tilde{\bar
d}_i,\tilde h_1,\tilde h_2,\tilde g,\tilde W,\tilde Z,\tilde \gamma}
\\[5mm] L_i,\bar E_i,Q_i,\bar U_i,\bar D_1,H_1,H_2,G, \cdots
\end{array}
$$
Thus $L_i(Q_i)$ and $\bar E_i(\bar U_i,\bar D_i)$ are the left-handed
lepton (quark) doublet and antilepton (antiquark) singlet chiral
superfields, where $i$ refers to the generation index.  Similarly
$H_{1,2}$ are the chiral superfields representing the two Higgs doublets
and $G$ the vector superfield representing the gluon.  The superpotential
contains the Higgs Yukawa coupling terms responsible for the lepton and
quark masses
$$
W_1 = h_{ij} L_i H_2 \bar E_j + h'_{ij} Q_i H_2 \bar D_j +
h^{\prime\prime}_{ij} Q_i H_1 \bar U_j.
\eqno (1)
$$
However, these are not the only Yukawa couplings allowed by the $SU(3)
\times SU(2) \times U(1)$ gauge invariance and supersymmetry.  They allow
three additional Yukawa coupling terms
$$
W_2 = \lambda_{ijk} L_i L_j \bar E_k + \lambda'_{ijk} L_i Q_j \bar D_k +
\lambda^{\prime\prime}_{ijk} \bar D_i \bar D_j \bar U_k.
\eqno (2)
$$
They can be rewritten in terms of the scalar and fermionic components of
the chiral superfield, i.e.
$$
W_2 = \lambda_{ijk} \ell_i \tilde\ell_j \bar e_k + \lambda'_{ijk} \ell_i
\tilde q_j \bar d_k + \lambda^{\prime\prime}_{ijk} \bar d_i \tilde{\bar d}_j
\bar u_k
\eqno (3)
$$
where the supertwiddle denotes the scalar partner of lepton and quark
(i.e. the slepton and squark), and there are analogous terms from the
permutation of the supertwiddle.  Evidently the 1st and 2nd terms of eq.
(3) violate lepton number (L) and the 3rd term violates baryon number (B)
conservation.  Note that there were no such couplings in the standard
model due to its particle content and Lorentz invariance; i.e. absence of
triple fermion coupling is simply ensured by angular momentum
conservation.  In the presence of scalar quarks and leptons, however,
there is no basic principle of physics which would prohibit these lepton
and baryon number violating couplings.  Evidently the source of lepton and
baryon number violation in the above couplings is the single emission
(absorption) of a superparticle as denoted by the supertwiddle.  Thus it
is jointly referred to as $R$-parity violation, where
$$
R = (-1)^{3B+L+2S}
\eqno (4)
$$
is so defined that it is $+1$ for all the standard model particles and
$-1$ for their superpartners, differing by $1/2$ units of spin(S).

\bigskip

\noindent {\bf Properties of the $R\!\!\!\!/$ Yukawa Couplings} : \\

\nobreak
Let us consider some general properties of the above $R\!\!\!\!/$ Yukawa
couplings.

\begin{enumerate}

\item[(i)] The $SU(2)$ and $SU(3)$ gauge invariance of the superpotential
require that the $\lambda_{ijk} L_i L_j \bar E_k$ and
$\lambda^{\prime\prime}_{ijk} \bar D_i \bar D_j \bar U_k$ terms are
antisymmetric combinations of the first two superfields in $SU(2)$ and
$SU(3)$ respectively.  Since the superfields satisfy Bose statistics, this
implies that the couplings are antisymmetric in the first two indices
$$
\lambda_{ijk} = -\lambda_{jik}, ~~~\lambda^{\prime\prime}_{ijk} =
-\lambda^{\prime\prime}_{jik}.
\eqno (5)
$$
Thus there are 9 independent $\lambda$ and $\lambda^{\prime\prime}$
coupling.  Together with the 27 $\lambda'$ couplings one has a total of 45
independent Yukawa couplings.

\item[(ii)] In analogy with the observed hierarchy of the Higgs Yukawa
couplings, as reflected by the quark and lepton masses, it is reasonable
to assume a hierachical structure for these additional Yukawa couplings as
well [2-4].  Thus one expects one of these 45 independent Yukawa couplings
to dominate over the others; but one does not know a priori which one.

\item[(iii)] Proton stability requires the $\lambda'$ or
$\lambda^{\prime\prime}$ coupling to be vanishingly small,
$$
\lambda' ~{\rm or}~ \lambda^{\prime\prime} \simeq 0 ~({\rm i.e.}~ \ll
10^{-10}).
\eqno (6)
$$
For these couplings would lead to proton decay via squark exchange as
shown in Fig. 1.  Since the SUSY solution to the gauge hierarchy problem
requires
$$
m_{\tilde q} \sim m_W \sim 100 ~{\rm GeV},
\eqno (7)
$$
this would imply a proton life time typical of the weak decay scale, i.e.
$\tau_P \sim 10^{-8} ~{\rm sec}$!  The experimental limit on the proton
life time, $\tau_P > 10^{32}~{\rm sec}$, implies the above constraint on
$\lambda'$ and/or $\lambda^{\prime\prime}$.
\end{enumerate}

\vskip 4cm
\begin{enumerate}
\item[Fig. 1.] Proton decay via squark exchange in $R\!\!\!\!/$ SUSY
model.
\end{enumerate}

\bigskip

\noindent {\bf $R$ Conserving SUSY Model and the Missing-$p_T$ Signature}
: \\

\nobreak
The traditional prescription for overcoming the proton decay problem has
been to banish all the $R\!\!\!\!/$ Yukawa couplings of eq. 2.  This
results in the standard $R$ conserving SUSY model.  In particular the $R$
conservation implies that (a) both baryon and lepton numbers are
conserved, (b) the superparticles are produced in pair and (c) all of them
decay into the lightest superparticle (LSP) which has to be stable.  The
stable LSP cannot carry any colour or electric charge for cosmological
reasons.  Thus the LSP can be either the sneutrino $\tilde \nu$ or the
photino $\tilde \gamma$ -- the latter being the case in most of the SUSY models
in the market.  Finally the photino interacts with ordinary matter as
weakly as the neutrino, as shown in Fig. 2 (the same is true for
sneutrino).

\newpage
\vspace*{5cm}
\begin{enumerate}
\item[Fig. 2.] Comparison of photino and neutrino interaction rates with
ordinary matter, i.e. electrons and quarks.
\end{enumerate}

Thus it would escape the detector without a trace like the
neutrino.  The resulting imbalance in visible momentum provides the
canonical missing transverse momentum signature for superparticle search
(momentum balancing in the longitudinal direction is not possible in a
collider due to loss of particles along the beam pipe).

The missing-$p_T$ signature has been extensively used in the superparticle
search at the $\bar pp$ colliders as well as LEP.  Ofcourse the
missing-$p_T$ signature is not very crucial for the LEP results [5], most
of which follow simply from the measurement of $Z$ total width, i.e.
$$
\Gamma_Z^{\rm Expt.} = \Gamma_Z^{SM} \Rightarrow m_{\tilde\ell,\tilde
q,\tilde W,\tilde h} > {1\over2} m_Z.
\eqno (8)
$$
The only superparticles escaping this mass limit are the neutral
gauginos ($\tilde \gamma,\tilde Z,\tilde g$), which do not couple to $Z$.  On
the other hand practically all the superparticle searches carried out at
the $\bar pp$ colliders so far rely heavily on the missing-$p_T$ signature
[6].  One expects strong production of squark and gluino pairs followed by
their prompt decay into the LSP ($\tilde \gamma$), as shown in Fig. 3.  The
escaping photinos provide the missing-$p_T$ signature for squark/gluino
production at the $\bar pp$ collider.

\newpage

\vspace*{4cm}
\begin{enumerate}
\item[Fig. 3.] Strong production of gluino and squark pair, followed by
their decay into the LSP $(\tilde \gamma)$.
\end{enumerate}
A recent analysis of the
missing-$p_T$ data from the Tevatron collider has given a squark/gluino
mass limit of [7]
$$
m_{\tilde q,\tilde g} \gsim 140~{\rm GeV},
\eqno (9)
$$
which represents the strongest experimental limit on any superparticle
mass.  But evidently it relies heavily on the assumption of $R$-parity
conservation.

\bigskip
\noindent {\bf $R$ Violating SUSY Model and the Multilepton Signature} :\\

\nobreak
The superparticle searches have been largely restricted so far to the
missing-$p_T$ channel due to the underlying assumption of $R$
conservation.  There is a growing realisation however that one should
remove these blinkers and look for possible superparticle signals outside
the missing-$p_T$ channel, since there is no compelling reason for $R$
conservation in the first place [4-6].  While $R$ conservation implies
proton stability the converse is not true -- i.e. proton stability implies
baryon or lepton number conservation, but not necessarily both.  In
particular it allows sizable values for either the $L\!\!\!\!/$ Yukawa
couplings $\lambda(\lambda')$ or the $B\!\!\!\!/$ couplings
$\lambda^{\prime\prime}$.  Consequently one has two types of $R\!\!\!\!/$
SUSY models consistent with proton stability, i.e.
$$
{\rm (a)} ~\lambda ~{\rm or}~ \lambda' \not= 0 ~(L\!\!\!\!/ ~{\rm Model})
\eqno (10a)
$$
$$
{\rm (b)}~ \lambda^{\prime\prime} \not= 0 ~(B\!\!\!\!/ ~{\rm Model}).
\eqno (10b)
$$
In either case the LSP is unstable, i.e. no missing-$p_T$ signature.
Assuming the LSP to be the lightest neutralino $(\tilde\wedge)$, one
expects the decays (Fig. 4)
$$
\tilde\wedge {\buildrel\tilde\ell \over \longrightarrow} \ell_i \ell_j
\bar e_k, ~~\tilde\wedge {\buildrel \tilde q \over \longrightarrow}
\ell_i q_j \bar d_k
\eqno (11a)
$$
or
$$
\tilde\wedge {\buildrel \tilde q \over \longrightarrow} d_i d_j u_k
\eqno (11b)
$$
depending on whether the dominant $R\!\!\!\!/$ Yukawa coupling is a
$\lambda_{ijk}, \lambda'_{ijk}$ or $\lambda^{\prime\prime}_{ijk}$ coupling.

\vspace*{4cm}
\begin{enumerate}
\item[Fig. 4.] $R\!\!\!\!/$ decay of the LSP, assumed to be the lightest
neutralino $\tilde\wedge$, where $f$ denotes either quark or lepton.
\end{enumerate}

To be more precise, the LSP decays within the detector ($\sim$ 1 meter) if
the dominant $R\!\!\!\!/$ coupling is
$$
\gsim 10^{-5} (m_{\tilde\ell,\tilde q}/100~{\rm GeV})^2
(m_{\tilde\wedge}/30~{\rm GeV})^{-5/2},
\eqno (12)
$$
below which it would decay outside simulating the missing-$p_T$ signature
[4].  While $B\!\!\!\!/$ decay of eq. 11b would correspond to a multijet
final state which is indistinguishable from the QCD background, the
$L\!\!\!\!/$ decays of eq. 11a would result in a distinctive multilepton
final state from the decay of the $\tilde\wedge$ pair.  More over
$\tilde\wedge$ being a Majorana particle, it decays into the final states
of eq. 11 as well as their charge conjugate states with equal probability.
This would lead to a distinctive final state with like sign dileptons.
Thus the multilepton channel provides a superparticle signature in the
$L\!\!\!\!/$ SUSY model which is as viable as the missing-$p_T$ signature
for the $R$ conserving case.  Indeed we shall see below that the Tevatron
dilepton data provides a squark/gluino mass limit for the $L\!\!\!\!/$
SUSY model, which is comparable to eq. 9 above.  Before discussing this
result, however, it will be useful to briefly review the theoretical and
phenomenological status of $R$ parity breaking SUSY models.

\bigskip

\noindent {\bf Theoretical Ideas on $R$ (Non) Conservation} :\\

\nobreak
There is no theoretical basis for any of the global symmetries, $B,L$ or
$R$, within the MSSM.  Thus the origin of proton stability lies outside
MSSM.  One hopes this to be ensured by a discrete symmetry arising from
the underlying string theory.  In this context it has been recently shown
by Ibanez and Ross [8] that there are only two such discrete symmetries
which are discrete anomaly free and consistent with the particle content
of MSSM.  They are the $Z_2$ and $Z_3$ symmetries corresponding to $R$ and
$B$ conservation respectively.  More over the latter has been shown to
have the advantage of eliminating the dimension 5 contribution to proton
decay along with the dimension 4 contribution of Fig. 1.  Thus from a
theoretical stand point the $L\!\!\!\!/$ SUSY model seems to be no less
attractive than the conventional $R$ conserving model.

\bigskip

\noindent {\bf Cosmological Constraint on $R\!\!\!\!/$ Couplings} :\\

\nobreak
GUT scale baryogenesis is expected to be washed away by a $B\!\!\!\!/$
SUSY interaction occuring at a lower energy scale of $\sim 100$ GeV.
Consequently the observed baryon asymmetry of the universe puts a severe
constraint [9] on the $B\!\!\!\!/$ Yukawa coupling of eq. 3, i.e.
$$
\lambda^{\prime\prime} < 10^{-7}.
\eqno (13)
$$
This means that if at all the $B\!\!\!\!/$ LSP decay occurs, it will be
outside the detector (eq. 12) and hence indistinguishable from the
missing-$p_T$ signature.  This is evidently a welcome result in the
absence of a distinctive signature for the $B\!\!\!\!/$ LSP decay.  It has been
further argued in [9] that the $L$ violating SUSY interaction can combine
with the $(B+L)$ violating nonperturbative electroweak interaction to wash
out any previously generated baryon asymmetry.  It should be noted however
that the latter interaction conserves not only $B-L$ but also $B/3 - L_i$
for each lepton generation [10].  Consequently the preservation of the
baryon asymmetry is ensured by the effective conservation of any one
lepton generation.  This implies the upperbound of eq. 13 for the smallest
$L\!\!\!\!/$ Yukawa couplings $\lambda_{ijk} (\lambda'_{ijk})$, while the
$L\!\!\!\!/$ LSP decay within the detector requires the lower bound of eq.
12 for the largest one.  With the expected hierarchy among these Yukawa
couplings it is evidently not difficult to satisfy both the requirements.
Note that the quark generations are not conserved unlike the leptons, so
that the upper bound of eq. 13 applies to all the indices of
$\lambda^{\prime\prime}_{ijk}$.  In summary, the observed baryon asymmetry
of the Universe seems to imply severe restrictions for the $B\!\!\!\!/$
LSP decay but not for the corresponding $L\!\!\!\!/$ decay [10].

\bigskip

\noindent {\bf Laboratory Constraints on $R\!\!\!\!/$ Couplings} :\\

\nobreak
Several phenomenological constraints on the $B\!\!\!\!/$ and $L\!\!\!\!/$
Yukawa couplings of eq. 3 have been obtained by considering virtual
superparticle exchange contributions to various processes, measured in the
laboratory.  For the $B\!\!\!\!/$ couplings there is only one serious
constraint following from the absence of $n - \bar n$ oscillation (Fig.
5a) and the corresponding heavy nuclei decay [11].  One gets
$$
\lambda^{\prime\prime}_{211} < 10^{-8} (m_{\tilde q}/100~{\rm GeV})^{5/2}.
\eqno (14)
$$
For the $L\!\!\!\!/$ Yukawa couplings, which are of greater interest to
us, the constraints are more numerous but much weaker than above.  The
strongest constraints follow form the radiative contribution to the
$\nu_e$ mass of Fig. 5b [2]
$$
\lambda_{133}, \lambda'_{133} < 10^{-2} (m_{\tilde \ell,\tilde q}/100~{\rm
GeV})^{1/2}
\eqno (15)
$$
and from the absence of neutrinoless double beta decay (Fig. 5c) [12]
$$
\lambda'_{111} < 10^{-2} (m_{\tilde \ell}/100~{\rm GeV})^{5/2}.
\eqno (16)
$$
There are weaker bounds from the observed charged current universality in
muon and neutron beta decays (Figs. 5d and e) [13]
$$
\lambda_{12k},\lambda'_{12k} < 0.04,0.03 (m_{\tilde \ell,\tilde
q}/100~{\rm GeV}).
\eqno (17)
$$
Under the assumption of hierarchy one of the two couplings dominates over
the other, so that each one can be constrained from the ratio of the two
decay rates.  There are constraints on several other $\lambda$ and
$\lambda'$ couplings; but they are still weaker that these ones [4].

Clearly none of the above bounds on $L\!\!\!\!/$ Yukawa couplings is
strong enough to inhibit the $L\!\!\!\!/$ LSP decay within the detector
(eq. 12).  Thus the multilepton channels are relevant for superparticle
search for a large part of the allowed coupling parameter space.

\vspace*{14cm}
\begin{enumerate}
\item[Fig. 5.] The $R\!\!\!\!/$ SUSY model contribution to (a)
neutron-antineutron oscillation, (b) $\nu_e$ mass, (c) neutrinoless double
beta decay,  (d) muon beta decay and (e) neutron beta decay.
\end{enumerate}

\newpage

\noindent {\bf Squark and Gluion Search in $R\!\!\!\!/$ SUSY Model with
the Tevatron Dilepton Data} :\\

\nobreak
We shall make two simplifying assumptions, leading to conservative mass
limits for the superparticles [14].

\begin{enumerate}

\item[1.] The largest $L\!\!\!\!/$ Yukawa coupling, responsible for the LSP
decay, is assumed to be significantly smaller than 1.  Thus we assume the
superparticles to be produced in pair and to decay into the leptonic
channel only via the LSP (Fig. 3).  It is clear that contributions from
single superparticle production and direct leptonic decay of squark via
eq. 3 will only enhance the multilepton signal and hence lead to a
stronger mass limit.

\item[2.] We shall explore gluino production by assuming it to be
significantly lighter than the squark and vice versa.  It is well known
that relaxing these constraints increases the signal and the
resulting mass limit [6,7].  Thus the squark (gluino) contribution to the
dilepton cross-section comes from the corresponding diagram of Fig. 3,
followed by the LSP decays into the dominant $L\!\!\!\!/$ channel of eq.
11a.  Consequently the squark (gluino) signal is independent of the gluino
(squark) mass.  It is also independent of the Yukawa coupling parameter
since the LSP decays wit  nearly 100\% branching ratio in to the dominant
$L\!\!\!\!/$ channel ($\ell_i\ell_j\bar e_k$ or $\ell_i q_j \bar d_k$)
with specific generation indices.  Only one has to take care of the
branching fractions into the 2 charge combinations of this channel,
corresponding to $\ell_i$ being a neutrino or a charged lepton.  They are
equal for $\ell_i\ell_j\bar e_k$; but depend on the nature of the LSP for
$\ell_i q_j \bar d_k$ [15].  One can see this by substituting the
corresponding Yukawa coupling terms
\end{enumerate}
$$
\begin{array}{l}
\lambda'\Bigg[\tilde e_L \bar d u_L + \tilde u_L \bar d e_L + \tilde{\bar
d}_L \bar e^c u_L \\[4mm]
{}~~~~- \tilde\nu_L \bar d d_L - \tilde d_L \bar d \nu_L - \tilde{\bar d}_L
\bar
d^c \nu_L\Bigg]
\end{array}
\eqno (18)
$$
into the LSP decay of Fig. 4.  In particular for photino decay one can see
that the relative branching fractions of the neutrino and charged lepton
channels are about 1:7.  For simplicity one generally assumes the relative
branching fractions to be equal [14,16], so that the branching fraction is
1/4 for the dilepton channel and 1/8 for the like sign dilepton.  Note
that the corresponding branching fractions for the above case would be 3/4
and 3/8, resulting in a larger dilepton cross-section.  We shall comment
on this latter.

Since the lepton momentum spectrum from the LSP decay is sensitive to the
LSP mass, a discussion of this assumption is in order.  While exploring
for gluino, we shall be interested in the mass range $m_{\tilde g} <
150~{\rm GeV}$.  With the MSSM mass relation [1,17]
$$
m_{\tilde \gamma} = {8\over3} {\alpha(M_Z) \over \alpha_S (M_Z)} m_{\tilde g}
\simeq {1\over5} m_{\tilde g}
\eqno (19)
$$
this corresponds to a photino mass range $m_{\tilde \gamma} < 30~{\rm GeV}$.
As we saw in eq. 8 above, LEP constraints essentially all other
superparticles to be heavier than 40 GeV [5].  Thus it is reasonable to
assume the LSP to be photino in this case with
$$
m_{\tilde \wedge} = m_{\tilde g}/5.
\eqno (20a)
$$
Indeed, this agrees with the explicit calculation of LSP mass and
composition incorporating the constraints of LEP and $m_{\tilde g} <
150~{\rm GeV}$ [17].  On the other hand, while exploring for squark over
this mass range we shall assume
$$
m_{\tilde\wedge} = 30~{\rm GeV}.
\eqno (20b)
$$
This corresponds to a conservative lower limit consistent with our
assumption of $m_{\tilde g}$ being significantly larger than $m_{\tilde
q}$ for this case.  However, this is adequate for our purpose, since a
higher LSP mass would correspond to a harder decay lepton momentum and
hence a larger signal.

The Tevatron dilepton ($ee$ or $\mu\mu$) data [18] can probe all those
$L\!\!\!\!/$ SUSY models where the dominant Yukawa coupling has a lepton
index 1 or 2.  Let us start by considering
$$
\lambda'_{1jk,2jk} ~({\rm or~equivalently}~\lambda_{133,233}),
$$
which leads to only one pair of electrons or muons and hence to the most
conservative dilepton signal.  Table 1 summarises the effect of various
experimental cuts on the signal cross-section [14].  The effect of the
lepton $p_T$ cut, shown in the 1st column, is quite strong because of the
sequencial decay.  It is ofcourse relatively stronger for the gluino due
to its 3-body decay.  The 2nd and 3rd columns refer to lepton isolation and
rapidity cuts.  The 4th and 5th columns refer to cuts on the dilepton
invariant mass and azimuthal angle to elliminate the $Z$ decay background.
The net detection efficiency is $\sim 1$\% for gluino and $\sim 3$\% for
squark signal.  Note that the last mentioned cut could be dispensed with
for the like sign dilepton channel.  Ofcourse the resulting gain $\sim 2$
in the acceptance factor will be compensated by a factor of 2 reduction in
the cross-section.

\bigskip
\bigskip
\bigskip
\oddsidemargin -10pt
\def \dc {\underline{D.~Choudhury}}
\def \us {U.~Sarkar}
\def \pr { {\it Phys. Rev.} }
\def \prl { {\it Phys. Rev. Lett} }
\def \pl { {\it Phys. Lett.} }
\def \mpl { {\it Mod. Phys. Lett.} }
\def\multic{ \multicolumn{1}{c|} }
\def\multir{ \multicolumn{1}{c||} }
\def\multil{ \multicolumn{2}{||c||} }
\begin{enumerate}
\item[Table 1 :] Acceptance factors for different kinematic cuts on
dilepton events.  (The net efficiency should be multiplied by a factor of
0.5 (0.7) for $ee (\mu\mu)$ events to take account of the geometrical
acceptance and $e(\mu)$ selection efficiency.)
\end{enumerate}
\noindent
\def\multic{ \multicolumn{1}{c|} }
\def\multir{ \multicolumn{1}{c||} }
\def\multil{ \multicolumn{2}{||c||} }
\def\scsty{\scriptstyle}

\hsize7in
            \begin{tabular}{||cr||c|c|c|c|c|l||}
            \hline
            \multil{$\scsty m_{\tilde g(\tilde q)}$ }& \multic{$\scsty
                     Trans.\ mom.$}
               & \multic{$\scsty Isolation$ }  & \multic{$\scsty Rapidity$}
                           &\multic{$\scsty Inv.\ mass$} & \multic{$\scsty
Azimu
   th.$}
                           &\multir{$\scsty Net$}
                 \\
            \multil{$\scsty in \ GeV$  }& \multic{$\scsty p^T_\ell >
15 \ GeV$}
               & \multic{$\scsty p^T_{ac} < 5 \ GeV$}  &
\multic{$\scsty |y_\ell| < 1$}
                           &\multic{$\scsty M_{\ell\ell} \not=$} &
\multic{$\scsty
\phi_{\ell\ell } =$}
                           &\multir{$\scsty efficiency$}
                 \\
                 &   &&&&\multic{$\scsty 75-105 \ GeV$}& \multic{$\scsty$
20-120$^\circ$} &
                                 \\
            \hline
$m_{\tilde g} = $ & 75 & 0.25 & 0.58 & 0.75 & 0.88 & 0.34 & 0.0035 \\
                  &100 & 0.36 & 0.66 & 0.78 & 0.85 & 0.44 & 0.013 \\
                  &150 & 0.53 & 0.72 & 0.80 & 0.82 & 0.48 & 0.037 \\
\hline
            &&&&&&&\\
$m_{\tilde q} = $ & 75 & 0.45 & 0.79 & 0.74 & 0.83 & 0.33 & 0.020 \\
                  &100 & 0.53 & 0.73 & 0.75 & 0.80 & 0.43 & 0.030 \\
                  &150 & 0.64 & 0.60 & 0.77 & 0.78 & 0.52 & 0.035 \\
\hline
\end{tabular}
\newpage

\vspace*{14cm}
\begin{enumerate}
\item[Fig. 6.] The visibile dimuon and dielectron cross-sections,
following the kinematic cuts of Table 1, are shown as functions of gluino
(squark) mass.  The right hand scale shows the expected number of events
for the CDF luminosity of 4.4 pb$^{-1}$.  The 95\% CL limits shown
correspond to 1 dimuon and 2 dielectron events remaining in the CDF data
after these kinematic cuts.
\end{enumerate}

The resulting dimuon and dielectron cross-sections are shown against the
gluino (squark) mass in Fig. 6 [14].  The right-hand scale shows the
expected number of events corresponding to the integrated luminosity of
4.4 pb$^{-1}$ of the CDF data [18].  The data contains only one dimuon and
two dilepton events in the above kinematic region.  The corresponding 95\%
CL limits are also shown in the figure.  The resulting lower bound on
squark and gluino masses are
$$
m_{\tilde q,\tilde g} > 100 ~{\rm GeV}
\eqno (21)
$$
for the dielectron and somewhat larger for the dimuon case.

As mentioned above, the dilepton cross-section of Fig. 6 can be regarded
as the like sign dilepton cross-section without the azimuthal cut.  It
seems one can be reasonably certain that there are no like sign dilepton
events in the above data over the entire range of the azimuthal angle.
The resulting 95\% CL limit of 3 events, corresponding to 0 candidate
events, is at least a factor of 2 lower than the predicted rate for
$m_{\tilde q,\tilde g} = 100~{\rm GeV}$.  This factor can take care of the
uncertainty in the predicted rate, arising largely from the QCD
parametrisation.

Finally, let us consider the cases where the dominant $L\!\!\!\!/$
coupling is one of the remaining $\lambda$s.  Evidently the higher lepton
multiplicity will lead to a higher visible dilepton cross-section.
Indeed, in view of the small detection efficiency for each lepton one
expects the dilepton cross-section to be roughly propertional to the
multiplicity of the appropriate lepton in each LSP decay.  More over, $e$
and $\mu$ detection efficiencies being similar, one can treat them jointly
by combining the $ee,\mu\mu$ and $e\mu$ channels (none of which has any
like sign dilepton events).  In this way one can relate the size of the
visible (like-sign) dilepton cross-section for each case to that of Fig. 6
and derieve the corresponding mass limits.  The results are summarised in
Table 2.  The Table also shows the above mentioned enhancement of the
dilepton branching fraction from 1/4 to 3/4, corresponding to the decay of
the photino pair via $\lambda'_{1jk,2jk}$, and the resulting increase of
the mass limit from 100 to 130 GeV.

\newpage

\def\multil{ \multicolumn{1}{||c|} }
\def\multic{ \multicolumn{1}{|c|} }
\def\multir{ \multicolumn{1}{|c||} }
\begin{enumerate}
\item[Table 2:] Relative size of the like-sign or total dilepton
cross-section for different choices of the leading Yukawa coupling and the
corresponding limit on gluino/squark masses.
\end{enumerate}
           \begin{center}
                   \begin{tabular}{||c|c|c||}
                   \hline
                   \multil{ Leading Yukawa } & \multic{ Relative size of } &
           \multir { Limit of }\\
                   \multil{ coupling } & \multic{ $\sigma_{(\mu\mu+ee+e\mu)}$ }
   &
                   \multir{ $m_{\tilde g,\tilde g}$ }\\
                   \hline
                   $\lambda'_{3jk,i3k}$ & 0 & --\\
                   $\lambda'_{ijk} (i,j \not= 3)$ & 1 & 100 GeV\\
                    "~~~~~~~ & 3 & 130 GeV\\
                   \hline
                   $\lambda_{133,233}$ & 1 & 100 GeV\\
                   123 & 4 & 140 GeV\\
                   311,322,312,321 & 9 & 160 GeV\\
                   121,122 & 16 & 175 GeV\\
                   \hline
\end{tabular}
\end{center}

One sees from Table 2 that for most of the choices of the dominant
$L\!\!\!\!/$ Yukawa coupling one gets a squark/gluino mass bound
comparable to that of the $R$-conserving SUSY model (eq. 9).  The least of
these bounds, 100 GeV, holds for all but two choices of the dominant
$L\!\!\!\!/$ coupling.  While this is effective for all values of this
coupling larger than eq. 12, the $R$-conserving bound coming from the
missing-$p_T$ channel is valid for the complimentary region of this
coupling parameter.  Combining the two, gives a lower squark/gluino mass
bound of 100 GeV, which is valid for all values of the $L\!\!\!\!/$ Yukawa
couplings\footnote{An important by product is a corresponding photino
mass bound of $m_{\tilde \gamma} > 20$ GeV.} with these two exceptions.  The
two exceptions correspond to the dominant $L\!\!\!\!/$ coupling being
$\lambda'_{3jk}$ or $\lambda'_{i3k}$.  The former corresponds to the $\tau
q\bar q$ decay of LSP and the latter to the $\nu b\bar q$ decay, since the
corresponding charged lepton decay is inhibited by the large top quark
mass.  It would be hard to probe these channels in a hadron collider since
the $\tau$ identification is dificult in a multijet environment and the
missing-$p_T$ resulting from the $\nu$ is degraded by the long decay
chain.  These channels may be probed at LEP II upto a squark/gluino mass
of 100 GeV [19].

In summary, for almost all choices of the dominant $L\!\!\!\!/$ Yukawa
coupling one gets a squark/gluino mass bound from the CDF dilepton data,
which is comparable to that obtained from their missing-$p_T$ data.
Combining the two gives a squark/gluino mass bound of $\sim 100$ GeV,
which is valid for all values of the corresponding $L\!\!\!\!/$ Yukawa
coupling parameters.

\bigskip
\noindent {\bf Spontaneously $R$ Violating SUSY Model} :\\

\nobreak
Finally, let me comment on the alternative type of $R\!\!\!\!/$ SUSY
model, where the $R$-parity is spontaneously broken via a vacum
expectation value of the sneutrino [20].  This is not phenomenologically
viable for the MSSM.\footnote{In the MSSM the $\tilde\nu$ vacum
expectation value generates $SU(2)$ doublet Majorons, which are ruled out
by the LEP measurement of $Z$ invisible width as well as by astrophysical
constraints.}  However, there are phenomenologically viable models of
spontaneous $R$-parity breaking involving nonminimal SUSY models with
singlet $\nu(\tilde\nu)$ [21].  In these models, the LSP can decay either
into the leptonic channels considered above or into a neutrino and
Majoron, which is indistinguishable from the missing-$p_T$ channel.  Thus
the above squark/gluino mass bound, obtained by combining these two
channels, should be valid for these models as well.

\bigskip
\noindent {\bf Acknowledgement} :\\

\nobreak
I gratefully acknowledge discussions with Herbi Dreiner, Manuel Drees,
Pran Nath, Pran Nath Pandita, Graham Ross and Probir Roy, which have helped
to clarify a number of points discussed above.

\newpage

\noindent \underbar{\bf References}\\

\begin{enumerate}
\item[1.] H.P. Nilles, Physics Reports C110, 1 (1984); H.E. Haber and G.L.
Kane, Physics Reports C117, 75 (1985).

\item[2.] S. Dimopoulos and L.J. Hall, Phys. Lett. B207, 210 (1987).

\item[3.] S. Dawson, Nucl. Phys. B261, 297 (1985); S. Dimopoulos et al.,
Phys. Rev. D41, 2099 (1990).

\item[4.] H. Dreiner and G.G. Ross, Nucl. Phys. B365, 597 (1991).

\item[5.] See e.g., ALEPH Collaboration: D. Decamp et al., Physics Reports
C216, 253 (1992).

\item[6.] See e.g. E. Reya and D.P. Roy, Phys. Lett. B166, 223 (1986); R.M.
Barnett, H.E. Haber and G.L. Kane, Nucl. Phys. B267, 625 (1986); UA1
Collaboration: C. Albajar et al., Phys. Lett. B198, 261 (1987).

\item[7.] CDF Collaboration: F. Abe et al., Phys. Rev. Lett. 69, 3439 (1992).

\item[8.] L.E. Ibanez and G.G. Ross, Nucl. Phys. B368, 3 (1992).

\item[9.] B.A. Campbell, S. Davidson, J. Ellis and K. Olive, Phys. Lett.
B256, 457 (1991); W. Fishler, G.F. Giudice, R.G. Leigh and S. Paban, Phys.
Lett. B258, 45 (1991).

\item[10.] H. Dreiner, Proc. LP-HEP Conference, Geneva, 1991 (World
Scientific Press).

\item[11.] F. Zwirner, Phys. Lett. B132, 103 (1983); R. Barbieri and A.
Masiero, Nucl. Phys. B267, 679 (1986).

\item[12.] R.N. Mahaptra, Phys. Rev. D34, 3457 (1986).

\item[13.] V. Barger, G.F. Giudice and T. Han, Phys. Rev. D40, 2987 (1989).

\item[14.] D.P. Roy, Phys. Lett. B283, 270 (1992).

\item[15.] J. Butterworth and H. Dreiner, Oxford Preprint, OUNP-92-15 (1992).

\item[16.] H. Dreiner and R.J.N. Phillips, Nucl. Phys. B367, 591 (1991).

\item[17.] J. Ellis, G. Ridolphi and F. Zwirner, Phys. Lett. B237, 423
(1990); L. Roszkowski, Phys. Lett. B252, 471 (1990) and B262, 59 (1991).

\item[18.] CDF Collaboration: G.P. Yeh, Proc. of Les Recontres de Physique de
la Vallee d'Aoste, Italy (1990); Fermilab-Conf-90/138-E (1990).

\item[19.] R.M. Godbole, Probir Roy and Xererxes Tata, TIFR/TH/92-29 (1992).

\item[20.] C.S. Aulakh and R.N. Mohapatra, Phys. Lett. B119, 136 (1982); C.G.
Ross and J.W.F. Valle, Phys. Lett. B151, 375 (1985).

\item[21.] R. Arnowitt and P. Nath, Proc. of Neutrino 88 (World Scientific,
1988); M.C. Gonzalez-Garcia, J.G. Ramao and J.W.F. Valle, Madison preprint
MAD/PH/678 (1991); R.N. Mohapatra and T.G. Rizzo, Maryland preprint
UMD-PP-93-90.

\end{enumerate}

\end{document}